\documentclass[aps,showpacs,floats,amsmath,amssymb]{revtex4}
\usepackage{graphics}
\usepackage[next]{inputenc}
\usepackage[dvips]{epsfig}
\def\be{\begin{equation}}
\def\ee{\end{equation}}

\def\bi{\begin{itemize}}
\def\ei{\end{itemize}}
\def\bn{\begin{enumerate}}
\def\en{\end{enumerate}}
\def\bea{\begin{eqnarray}}
\def\eea{\end{eqnarray}}
\def\no{\nonumber}
\def\ba{\begin{array}}
\def\ea{\end{array}}
\def\bd{\begin{displaymath}}
\def\ed{\end{displaymath}}

\begin{document}
\title
{Second order quantum renormalisation group of $XXZ$ chain with next nearest neighbour interactions}
\author{R. Jafari$^{1}$ and A. Langari$^{1,2}$
\footnote{Correponding author: A. Langari, Institute for Advanced Studies in Basic Sciences, P. O. Box. 45195-1159, Zanjan, Iran.
e-mail: langari@cpfs.mpg.de, }}
\address{$^{1}$Institute for Advanced Studies in Basic Sciences, 45195-1159,
Zanjan, Iran\\
$^{2}$Max-Planck-Institut f\"ur Chemische Physik fester Stoffe,
01187 Dresden, Germany}
\date{\today}
\begin{abstract}

We have extended the application of quantum renormalisation group
(QRG) to the anisotropic Heisenberg model with next-nearest
neighbour (n-n-n) interaction. The second order correction has to be
taken into account to get a self similar renormalized Hamiltonian  
in the presence of n-n-n-interaction.
We have obtained the phase diagram of this model which consists of
three different phases, i.e, spin-fluid, dimerised and N\'{e}el
types which merge at the tri-critical point.
The anisotropy of the n-n-n-term changes the phase diagram
significantly. It has a dominant role in the N\'{e}el-dimer phase boundary.
The staggered magnetisation as an order parameter defines
the border between fluid-N\'{e}el and N\'{e}el-dimer phases. 
The improvement of the second order RG corrections on the ground state energy
of the Heisenberg model is presented. 
Moreover, the application of second order QRG on the spin lattice
model has been discussed generally. 
Our scheme shows that higher order corrections lead to an effective Hamiltonian with infinite range
of interactions.

\leftskip 2cm \rightskip 2cm
\end{abstract}
\pacs{75.10.Jm, 75.10.Pq, 75.40.Cx}

\maketitle
{\it Keywords: Quantum renormalization group, Spin chains, Heisenberg model, Quantum phase transitions,\\ Phase diagram, Second order corrections}

\section{Introduction}
Quantum phase transition has been one of the most interesting
topics in the area of strongly correlated systems in the last
decade. It is a phase transition at zero temperature where the
quantum fluctuations play the dominant role \cite{vojta}.
Suppression of the thermal fluctuations at zero temperature
introduces the ground state as the representative of the system.
The properties of the ground state may be changed drastically shown as a
non-analytic behaviour of a physical quantity by reaching the
quantum critical point. This can be done by tuning a parameter in
the Hamiltonian, for instance the magnetic field or the amount of
disorder. The study of the ground state and its energy is thus of
central importance for understanding the critical behaviour of
such systems.

The technique of renormalisation group (RG) has been so devised to
deal with these multi-scale problems \cite{Wilson,Drell,Stella}.
In the momentum space RG which is suitable for studying the
continuous systems, one iteratively integrates out small scale
fluctuations and renormalizes the Hamiltonian. In the real space RG,
which is usually performed on the lattice systems with discrete
variables (i.e quantum spin chain), an original
Hamiltonian is replaced with an effective one for a lower energy
subspace iteratively. In this approach the Hamiltonian is divided into inter-block ($H^{BB}$)
and intra-block parts ($H^{B}$), $H^{B}$ is diagonalized exactly and then $H^{BB}$
is projected into the low energy subspace of $H^{B}$ \cite{miguel1}.
The accuracy of this
method is determined by the number of states kept in the $H^{B}$ subspace and 
the approach to consider the effect of neglected subspace.
The Ising model in a transverse field
\cite{Drzewinski} and the anisotropic Heisenberg model
\cite{langari98} have been studied by quantum renormalisation
group (QRG) approach  which gives the correct phase diagram.
Moreover, the recent study on a more general model, $XYZ$ in a
transverse field, supports the power of this method to study the
collective behaviour of the spin models \cite{Langari2004}.

In this paper we are going to study the effect of higher order
corrections on the QRG scheme.
In this respect we will consider the one dimensional
$S=\frac{1}{2}$ antiferromagnetic $XXZ$ chain with next-nearest
neighbour (n-n-n) interactions. Because, even in the case of nearest 
neighbour Heisenberg model the n-n-n interaction will be generated 
for the renormalized Hamiltonian if we add the second order corrections.
Moreover, we will obtain the phase diagram of this model which is a function 
of the anisotropies and the n-n-n interactions.
We have calculated the effective Hamiltonian up to second order in
$H^{BB}$. 
The second order correction improves the accuracy of the results.
However, it must be taken into account to get
a self similar Hamiltonian after each step of QRG for the n-n-n $XXZ$ chain. 
In this
approach we have considered the effect of the whole states of the
block Hamiltonian which are partially ignored in the first order
approach. The present scheme allows us to have the analytic RG
equations, which gives a better understanding of the behaviour of
system by running of the coupling constants. We have succeeded to
obtain the phase diagram which contains the critical surface between
the spin-fluid, N\'{e}el and dimer phases. The boundaries between these phases
merge at the tri-critical point.
The projection of our phase diagram on the $\Delta=\delta$ plane
(the same anisotropy for nearest and next-nearest neighbour interactions)  
is in good agreement with the numerical ones \cite{Nomura}.
The QRG equations show that the anisotropy in the n-n-n term changes the
phase boundaries significantly. 
We have also derived the
staggered magnetisation in the z direction, that is the proper
order parameter to show the phase transition between
N\'{e}el-dimer and fluid-N\'{e}el phases. 

In the recent years,
several interesting quasi-one-dimensional magnetic systems have
been studied experimentally \cite{Hase,Motoyama,Coldea}. Among
them, some compounds containing $CuO$ chains with edge-sharing
$CuO_{4}$ plaquette were expected to be described by the $XXZ$
model with next-nearest neighbour interaction \cite{Mizuno}. The
Hamiltonian of this model on a periodic chain of $N$ sites is:
\be \label{eq1} 
\hspace{-2cm}H=\frac{J}{4}
\Big\{ \sum_{i=1}^{N}[\sigma_{i}^{x}\sigma_{i+1}^{x}+
\sigma_{i}^{y}\sigma_{i+1}^{y}+
\Delta~\sigma_{i}^{z}\sigma_{i+1}^{z}
+J_{2}(\sigma_{i}^{x}\sigma_{i+2}^{x}+
\sigma_{i}^{y}\sigma_{i+2}^{y}+
\delta~\sigma_{i}^{z}\sigma_{i+2}^{z})]\Big\},
\ee
where $J>0$ and $J\times J_{2}\geq0$ are the first and second-nearest
neighbour exchange couplings and the corresponding easy-axis
anisotropies are defined by $\Delta$ and $\delta$. For
$J_{2}=0$, the ground state properties are well known by the
Bethe ansatz \cite{Cloizeaux} and the bosonization technique. In
this case, for $\Delta<-1$, the system is in the ferromagnetic
phase, while for $\Delta>1$, it enters the N\'{e}el phase where
the twofold degenerate ground states are separated from the
excited ones by a finite gap \cite{Yang}. In the case of $XY$-type
anisotropy ($-1\leq\Delta\leq 1$) quantum fluctuations destroy the
long range order even at the zero temperature. The ground state is
characterised by  gapless excitations and algebraic decay of spin
correlations (spin-fluid). A kind of frustration can be
introduced to this model by adding the next-nearest
neighbour interaction ($J_{2}\neq0$) as well as the nearest
neighbour ones. On the $J_{2}=\frac{1}{2}, \Delta=\delta=1$ line, 
the model is known as the Majumdar-Ghosh Hamiltonian where the exact ground state
has been obtained to be purely dimerised  \cite{Majumdar}. 
The dimer state is characterised by the
excitation gap, the exponential decay of the spin correlation
functions and the dimer long-range order. According to the above
facts, there should exist the fluid, dimer and  N\'{e}el phases
in this model.
The isotropic model have been studied intensively by analytic and numerical
approaches \cite{Majumdar}--\cite{Gerhardt}.
However, the QRG approach to this model
has not been considered yet which gives a clear phase diagram 
in the presence of anisotropies. Moreover, as mentioned earlier, we are
going to address the implementation of higher order RG corrections and
discuss its features.
We will explain the QRG scheme in the next section where the
second order effective Hamiltonian and the renormalisation of
the coupling constants are obtained. In Sec. III , We will  present
the phase diagram and explain the border between different phases. 
Finally, we will discuss on the features of the second order approach and higher order corrections.


\section{RG Equations \label{sec2}}
The main idea of the RG method is the mode elimination or thinning
of the degrees of freedom followed by an iteration which reduces
the number of variables step by step until reaching a fixed point.
We have implemented the Kadanoff's block
method for this purpose, because they are well suited to perform
analytical calculations in the lattice models and they are
conceptually easy to be extended to the higher dimensions. In
Kadanoff's method, the lattice is divided into $N'$ blocks with
$n_{s}$ sites each  and the original non-diagonal Hamiltonian is
replaced with an effective one for a lower energy subspace. The
Hamiltonian can be written 
\bea
\label{eq2}
H=H^{B}+\lambda H^{BB},
\eea
where the block Hamiltonian $H^{B}$ is a sum of commuting
Hamiltonians, each acting on every block and $\lambda$ is a
coupling constant which is already in $H$ or else it can be
introduced as a parameter characterising the inter-block coupling
which can be set to 1 at the end of calculation. An exact
implementation of this method \cite{miguel1} is given by the
following equation,
\bea
\label{eq3}
H^{eff}=P_{B}HP_{B},
\eea
where $P_{B}$ is the projection operator. Eq.(\ref{eq2}) suggests
that we should search for the solution of the Eq.(\ref{eq3}) in
the form of a perturbative expansion in the inter-block coupling
parameter $\lambda$, namely
\bea
\label{eq4}
P_{B}=P_{0}+\lambda
P_{1}+\lambda^{2}P_{2}+\cdots~,~ H^{eff}=H^{eff}_{0}+\lambda
H^{eff}_{1}+\lambda^{2}H^{eff}_{2}+\cdots.
\eea
To the zeroth order in $\lambda$ Eq.(\ref{eq3}) becomes
\bea
\label{eq5}
H^{eff}_{0}=P_{0}H^{B}P_{0}.
\eea
Since $H^{B}$ is a sum of disconnected block Hamiltonians
\bea
\no
H^{B}=\sum_{I=1}^{N'}h_{I}^{B},
\eea
one can search for a solution of $P_{0}$ in a factorised form
\bea
\no
P_{0}=\prod_{I=1}^{N'}P_{0}^{I}.
\eea
It can be found that
\bea
\no
P_{0}^{I}=\sum_{i=1}^{k}|\psi_{i}\rangle\langle\psi_{i}|,
\eea
where $|\psi_{i}\rangle~(i=1,\cdots, k)$ are the $k$ lowest energy
states of $h_{I}^{B}$. First and second order corrections are
obtained to be
\bea
\label{eq6}
H_{1}^{eff}=P_{0}H^{BB}P_{0}~~~,~~~H_2^{eff}=P_{0}H^{BB}P_{1},
\eea
where $P_{1}$ is defined by the following equation
\bea
\no
P_{1}=(1-P_{0})\frac{1}{E_{0}-H^{B}}(1-P_{0})H^{BB}P_{0}.
\eea

\begin{figure}
\begin{center}
\includegraphics[width=12cm]{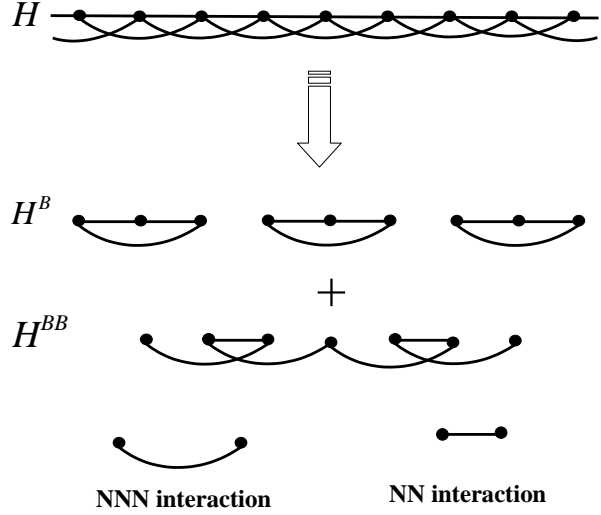}
\caption{The decomposition of chain into the three site blocks
Hamiltonian ($H^{B}$)
and the inter-block Hamiltonian ($H^{BB}$).}
\label{fig1}
\end{center}
\end{figure}

The second order correction can be written\cite{miguel-book,miguel-2},
\bea
\label{eq7}
H_2^{eff}=P_{0}[H^{BB}(1-P_{0})\frac{1}{E_{0}-H^{B}}(1-P_{0})H^{BB}]P_{0}.
\eea
We have considered three-site block (Fig.(\ref{fig1})) and kept
the degenerate ground states ($|\psi_{0}\rangle, |\psi_{0}'\rangle$) of each
block to construct the projection operator
($P_{0}=|\psi_{0}\rangle\langle\psi_{0}|+|\psi_{0}'\rangle\langle\psi_{0}'|$).
The ground states and their corresponding eigenvalues of each block
are
\bea
\no
|\psi_{0}\rangle=\frac{1}{\sqrt{2+q^{2}}}(|\uparrow\uparrow\downarrow\rangle+
q|\uparrow\downarrow\uparrow\rangle+|\downarrow\uparrow\uparrow\rangle),\\
\no
|\psi_{0}'\rangle=\frac{1}{\sqrt{2+q^{2}}}(|\uparrow\downarrow\downarrow\rangle+
q|\downarrow\uparrow\downarrow\rangle+|\downarrow\downarrow\uparrow\rangle),\\
\no
e_{0}=\frac{J}{4}[J_{2}-\Delta-\sqrt{(J_{2}+\Delta-\Delta_{2})^{2}+8}],
\eea
where $\Delta_2=J_2 \delta$.
In this notation $|\uparrow\rangle$, $|\downarrow\rangle$ are the
$\sigma^{z}$ eigenbases, and
\bea
\no
q=-\frac{1}{2}(J_{2}+\Delta-\Delta_{2}+\sqrt{(J_{2}+\Delta-\Delta_{2})^{2}+8}).
\eea
The interaction between blocks defines the effective interaction
of the renormalised chain where each block is considered as a new
single site. Calculating the effective Hamiltonian to the first
order correction leads to the $XXZ$ chain without next nearest
neighbour interaction ($J'_{2}=0$), i.e the effective Hamiltonian
is not exactly similar to the initial one. The
next-nearest neighbour interaction is the result of the second
order correction. When the second
order correction is added to the effective Hamiltonian, the
renormalised Hamiltonian apart from an additive constant ($e_{B}$)
is similar to Eq.(\ref{eq1}) with the renormalised couplings. Thus,
the effective Hamiltonian including  the second order correction is
\bea
\no
H^{eff}=\frac{J'}{4}&&\left[\sum_{i}^{N/3}({\sigma}_{i}^{x}{\sigma}_{i+1}^{x}+{\sigma}_{i}^{y}{\sigma}_{i+1}^{y})+
\Delta'({\sigma}_{i}^{z}{\sigma}_{i+1}^{z})\right.\\
\no
&&+\left.\sum_{i}^{N/3}J_{2}'({\sigma}_{i}^{x}{\sigma}_{i+2}^{x}+
{\sigma}_{i}^{y}{\sigma}_{i+2}^{y})+\Delta_{2}'({\sigma}_{i}^{z}{\sigma}_{i+2}^{z})\right]
+\frac{N}{3}e_{B}. \eea
The renormalised coupling constants are functions of the original
ones which are given by the following equations.
\bea
\label{eq8}
J'&&=J(\frac{2}{2+q^{2}})^{2}(~q^{2}+2J_{2}q~)-\frac{J^{2}}{4}(~\frac{\Delta}{e_{0}-e_{2}}~)(~\frac{q}{2+q^{2}}~)^{2}\\
\no
&&-\frac{J^{2}}{4}(~\frac{1}{e_{0}-e_{1}}~)(~\frac{1}{(~2+q^{2}~)(~2+p^{2}~)}~)^{2}(~p+q~)\times\\
\no
&&~~~~~~~~(~pq~)(~p+q+4J_{2}~)(~(~\Delta-2\Delta_{2}~)pq+4\Delta_{2}~)\\
\no
&&-\frac{J^{2}}{4}(~\frac{4}{2e_{0}-e_{1}-e_{2}}~)(~\frac{1}{2+q^{2}}~)^{2}(~\frac{q}{2+p^{2}}~)\times\\
\no
&&~~~~~~~~(~p+q+2J_{2}~)(~(~\Delta-\Delta_{2}~)pq+2\Delta_{2}~).
\eea

\bea
\label{eq9}
\Delta'=\left\{~J(~\frac{q}{2+q^{2}}~)^{2}(~\Delta
q^{2}+2\Delta_{2}(~2-q^{2}~)~)\right.\\
\no
~~~~~~~~-\frac{J^{2}}{4}(~\frac{1}{e_{0}-e_{1}}~)(~\frac{(~p+q~)^{2}+4J_{2}(~p+q~)}{(~2+q^{2}~)(~2+p^{2}~)}~)^{2}\\
\no
~~~~~~~~-\frac{J^{2}}{4}(~\frac{1}{e_{0}-e_{2}}~)(~\frac{q^{2}}{2(2+q^{2}~)}~)^{2}
-\frac{J^{2}}{4}(~\frac{1}{e_{0}-e_{3}}~)(~\frac{1+2J_{2}q}{2+q^{2}}~)^{2}\\
\no
~~~~~~~~-\frac{J^{2}}{4}(~\frac{2}{2e_{0}-e_{1}-e_{2}}~)(~\frac{1}{2+p^{2}}~)(~\frac{q(~p+q+2J_{2}~)}{2+q^{2}}~)^{2}\\
\no
~~~~~~~~+\frac{J^{2}}{4}(~\frac{2}{2e_{0}-e_{2}-e_{3}}~)(~\frac{q(~1+qJ_{2}~)}{2+q^{2}}~)^{2}\\
\no
~~~~~~~~+\left.\frac{J^{2}}{4}(~\frac{4}{2e_{0}-e_{1}-e_{3}}~)(~\frac{J_{2}(pq+q^{2}+2)+p+q}{(2+q^{2})(2+p^2)^{1/2}}~)^{2}~\right\}~/~J'.
\eea

\bea
\label{eq10}
J'_{2}=\left\{~\frac{J^{2}}{4}(~\frac{2}{2+q^{2}}~)^3\left[~\frac{(~J_{2}(~3q+p~)+q(~p+q~)~)^{2}}{(~e_{0}-e_{1}~)(~2+p^{2}~)}\right.\right.\\
\no
~~~~~~~~~~~~~~~~~~~~~~~~~~~~~+\left.\left.\frac{(~J_{2}(~1+q^{2}~)+q~)^{2}}{e_{0}-e_{3}}-\frac{(~q^{2}+J_{2}q~)^{2}}{2(~e_{0}-e_{2}~)}~\right]~\right\}~/~J'.
\eea

\bea \label{eq11}
\Delta'_{2}=\left\{~\frac{J^{2}}{4}(~\frac{2}{2+q^{2}}~)^{3}\left[~\frac{(~\Delta_{2}q(~p-pq^{2}~+q~)
+\frac{\Delta}{2}pq^{3}~)^{2}}{(~e_{0}-e_{1}~)(~2+p^{2}~)}\right.\right.\\
\no ~~~~~~~~~~~~~~~~~~~~~~~~~~~~~~~~-\left.\left.\frac{(~\Delta
q^{2}+\Delta_{2}(~2-q^{2}~)~)^{2}}{2(~e_{0}-e_{2}~)}\right]\right\}~/~J'.
\eea

\bea
\label{ea12}
e_{B}&&=\frac{1}{3}e_{0}\\
\no
&&+\frac{1}{3}\frac{J^{2}}{16}\left\{~(~\frac{1}{e_{0}-e_{1}}~)(~\frac{1}{(2+q^{2})(2+p^{2})}~)^{2}
(~(~(~p+q~)^{2}+4J_{2}(~p+q~)~)^{2}\right.\\
\no
&&+\frac{1}{2}(~p q(~\Delta+2\Delta_{2}(~2-p q~)~)~)^{2}~)
+(~\frac{1}{e_{0}-e_{2}}~)(~\frac{1}{2(~2+q^{2}~)}~)^{2}(~q^{2}+8\Delta^{2}~)\\
\no
&&+(~\frac{1}{e_{0}-e_{3}}~)(~\frac{~1+2J_{2}q}{2+q^{2}}~)^{2}
+(~\frac{2}{2e_{0}-e_{1}-e_{2}}~)(~\frac{1}{2+q^{2}}~)^{2}(~\frac{1}{2+p^{2}}~)\times\\
\no
&&~~~(~(~(p+q)q+2J_{2}q)^{2}+2(\Delta pq+\Delta_{2}(2-pq)~)^{2}~)\\
\no
&&+(~\frac{4}{2e_{0}-e_{1}-e_{3}}~)(~\frac{1}{2+q^{2}}~)^{2}(~\frac{1}{2+p^{2}}~)
(~(~p+q~)+J_{2}q(~p+q~)+2J_{2}~)^{2}\\
\no
&&+\left.(~\frac{2}{2e_{0}-e_{2}-e_{3}}~)
(~\frac{q}{2+q^{2}}~)^{2}(~1+J_{2}q~)^2~\right\}\\
\no
&&+(\frac{1}{3}\frac{J^{2}}{16})\left\{~(~\frac{2}{e_{0}-e_{3}}~)(~\frac{2}{2+q^{2}}~)^{3}(~q+J_{2}+J_{2}q^{2}~)^{2}\right.\\
\no
&&+\left.(~\frac{4}{e_{0}-e_{2}}~)(~\frac{1}{2+q^{2}}~)^{3}
(~2(~q^{2}+J_{2}q~)^{2}+(~\Delta~ q^{2}+2\Delta_{2}-\Delta_{2}~q^{2}~)^{2}~)\right.\\
\no
&&+\left.(~\frac{2}{e_{0}-e_{1}}~)(~\frac{1}{2+q^{2}}~)^{3}(\frac{1}{2+p^{2}})
\Big(8[~q^{2}+p~q+3J_{2}~q+J_{2}~p~]^{2}\right.\\
\no
&&+\left.[~p~q(~\Delta~q^{2}+2\Delta_{2}-\Delta_{2}~q^{2}~)+\Delta_{2}~q^{2}(~2-p~q~)~]^{2}\Big)~\right\}.
\eea
in which
\bea
\no
p=\frac{-1}{2}[~J_{2}+\Delta-\Delta_{2}-\sqrt{{(J_{2}+\Delta-\Delta_{2})}^{2}+8}~~].
\eea
\begin{figure}
\begin{center}
\includegraphics[width=14cm, angle=0]{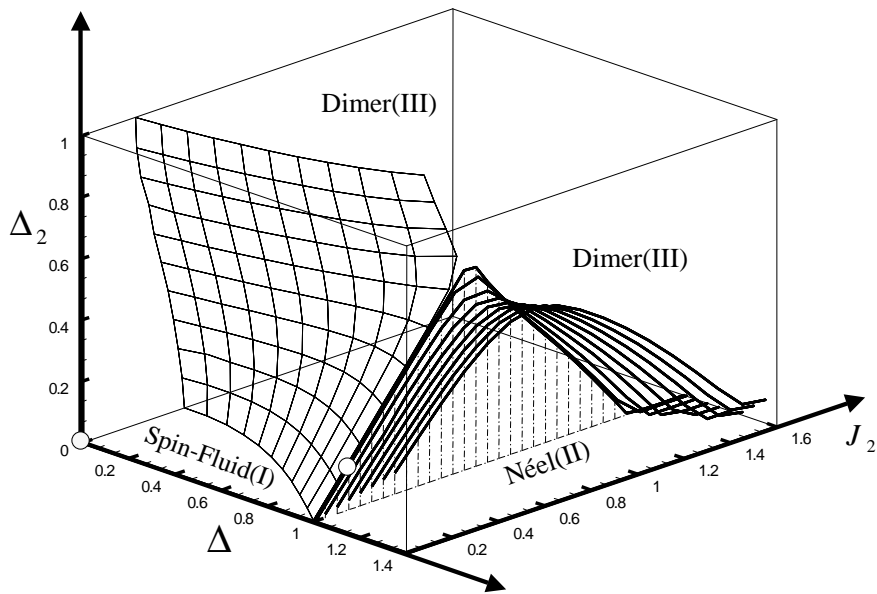}
\caption{
Phase diagram of the $XXZ$ model with next-nearest neighbour
interaction and different anisotropies. 
Open circles show the fixed points, the tri-critical point ($\Delta_2^*=J_{2}^*\simeq0.155, \Delta^*=1$)
and the XY fixed point ($\Delta_2^*=J_{2}^*= \Delta^*=0$) . For $0\leq\Delta\leq1$, the front
side of diagram is the spin-fluid phase (I) and the back side is 
the dimer phase (III). The checkerboard pattern
represents the boundary between them. In the case of $\Delta >1$ there are
two phases, the N\'{e}el-phase (II) which exists below the semi-circular lines and the dimer-phase (III) which
is above them. The boundary between N\'{e}el and dimer phases is shown by the semi-circular
lines. The surface which is depicted by the dash-dotted lines is the boundary between the dimer-N\'{e}el phases which is a vertical plane at $\Delta=1$. 
The phase transition between the spin-fluid phase (I) and the dimer phase (III)
occurs at the vertical plane $\Delta=1$ which has not been plotted in this
diagram to avoid it being complex.
}
\label{fig2}
\end{center}
\end{figure}

\begin{figure}
\begin{center}
\includegraphics[width=12cm]{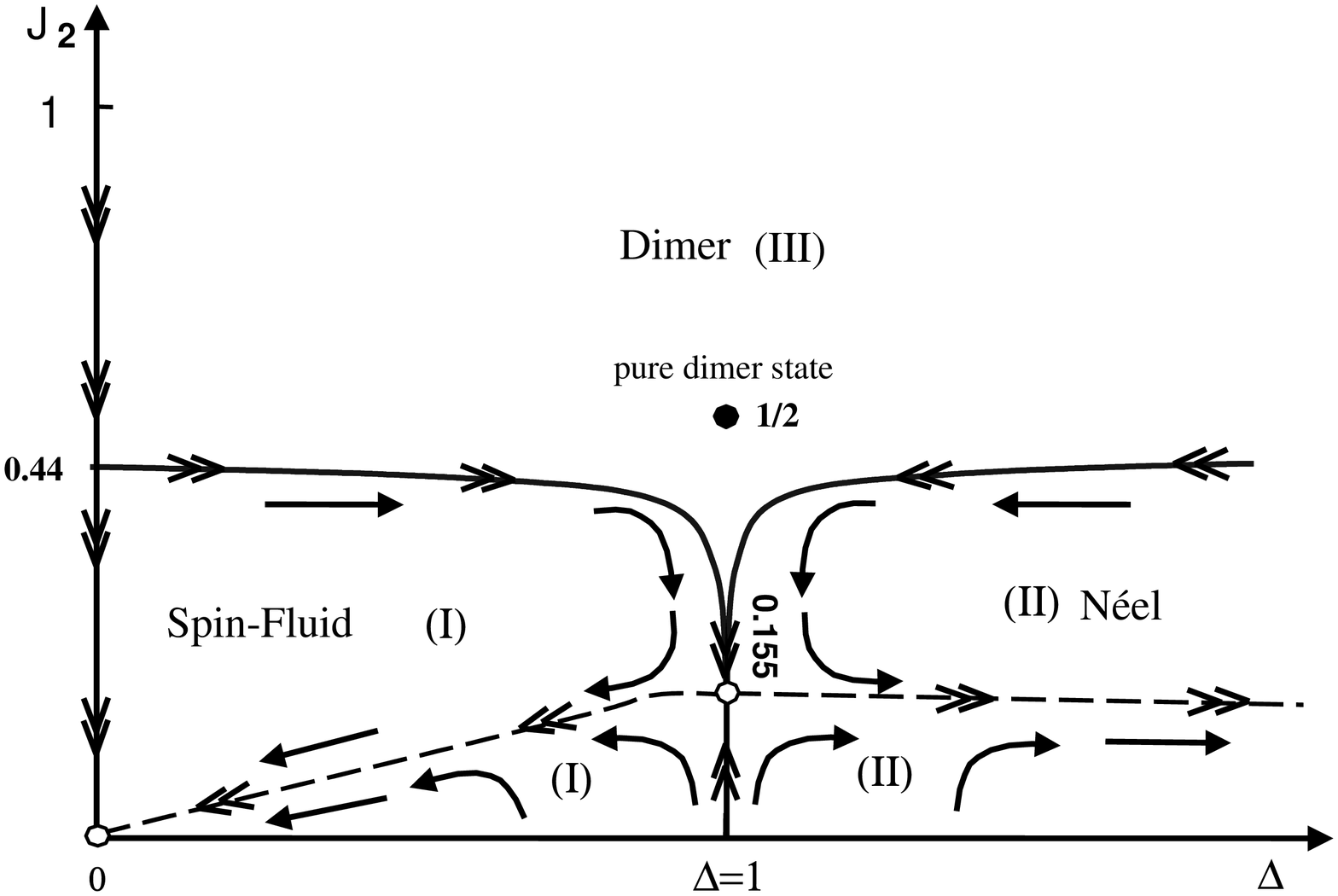}
\caption{The projection of the 3-dimensional phase diagram (Fig.(\ref{fig2})) on 
the $\Delta_2=J_2 \Delta$ plane.
Arrows show the running of couplings
under RG. Open circles show fixed points. The dashed lines in
the spin-fluid and N\'{e}el phases are the lines of the
Gaussian fixed point. Different phases are labeled by: (I)
spin-fluid, (II) N\'{e}el order, (III) dimer. The Filled circle denoted by $J_2=1/2$ on
the $\Delta=1$ line is the pure dimer state.} 
\label{fig3}
\end{center}
\end{figure}

In the above equations, $e_{1}, e_{2}$ and $e_{3}$ are the first,
second and third eigenenergies of the block's Hamiltonian which have
analytic expressions in terms of the coupling constants but not presented here.
Due to the level crossing which occurs for the
eigenstates of the block Hamiltonian, RG equations are valid for
$\frac{1}{2}[(\Delta-\Delta_2)-\sqrt{(\Delta-\Delta_2)^2+4}]<J_{2}<
\frac{1}{2}[(\Delta-\Delta_2)+\sqrt{(\Delta-\Delta_2)^2+4}]$. 
We have plotted the RG flow and
different phases in Fig.(\ref{fig2}) which will be discussed in
the next section.

\section{Phase diagram}
The RG equations show the running of $J$ coupling to zero which represents the
renormalisation of energy scale. At the zero temperature, phase
transition occurs upon variation of the parameters in the
Hamiltonian. In the region of planar anisotropy $0\leq\Delta<1$,
the nearest neighbour interaction ($J_{2}=0$) is known not to
support any kind of long range order and the ground state is the
so called spin-fluid state. Increasing the amount of anisotropy is
necessary to stabilize the spin alignment. For $\Delta>1$ the
ground state is the N\'{e}el ordered state. In the case of
$J_{2}>0$, the nearest neighbour and next-nearest neighbour
couplings are in competition with each other. The latter thus
frustrates the ordering tendency of the former. For
$0\leq\Delta<1$ the interplay of the two competing terms in the
presence of quantum fluctuations produces the dimer phase for
$J_{2}\geq J_{2}^{c}(\Delta, \Delta_2)$. 
Our RG equations show that the phase boundary between the  dimer
and the spin-fluid phase depends on the n-n-n anisotropy 
coupling ($\Delta_2$) and can be described by a two dimensional surface
which is convex in a view from the spin-fluid phase (see Fig.(\ref{fig2}),
the checkerboard curved plane).

The dimer or spin Peierls phase has
a spin gap and a broken translation symmetry (the unit cell is
doubled) in the thermodynamic limit. 
The dimer-fluid
transition is known to be of Berzinskii-Kosterlitz-Thouless (BKT)
type \cite{Berzinskii}. At the BKT transition point, the
divergence of the correlation length is not of the usual power law
type but very singular \cite{Kosterlitz}. In fact, all the
derivatives of the inverse of the correlation length at the
critical point are zero. Moreover, at the BKT critical point there
appear the logarithmic correction as a finite size effect which
converges very slowly, in various quantities, such as correlation
function and susceptibilities, therefore it is very difficult to
find the critical point of the BKT type transition accurately. 
Two effective methods have been introduced to find this critical point. 
In the work of Roomany et. al
\cite{Rommany},  $\beta$-function is calculated
numerically, while in the Nomura-Okamoto approach \cite{Nomura} 
the fluid-dimer transition is determined by the degeneracy between
the doublet excitation and the dimer excitation . Moreover, in
Ref.\cite{Gerhardt}  a proper structure factor has been introduced
to probe this transition. However, we determine the fluid-dimer phase
transition by using the running of couplings under RG. In the
spin-fluid phase, the anisotropy and next-nearest neighbour
couplings are irrelevant while in the dimer phase they run to the
tri-critical point ($\Delta_2^*=J_{2}^*\simeq0.155, \Delta^*=1$). 
For small $\Delta_2$ the fluid-dimer phase boundary ($J_{2}^{c}(\Delta)$) shows an
inclination to the lower values versus $\Delta$ while it behaves 
conversely for higher $\Delta_2$ as shown in Fig.(\ref{fig2}).
The comparison with numerical results \cite{Nomura}  for $\Delta=\delta$
shows  good qualitative agreement. 
The projection of the three dimensional phase diagram (Fig.(\ref{fig2})) on the $\Delta_2=J_2 \Delta$
plane is shown in Fig.(\ref{fig3}). Different phases and the running of RG flows 
can be observed simply.
However, from quantitative
point of view we got higher values for $J_{2}^{c}$ than the numerical results.
For instance, at $\Delta=0$ the RG analysis gives $J_{2}^{c}\simeq0.44$
which can be compared with the numerical result of $J_{2}^{c}\simeq0.33$
presented in Ref.\cite{Nomura}. The difference is inherited to the QRG
scheme where the boundary condition of the isolated block does not
represent the presence of the rest of chain \cite{noak,sierra}.
\begin{figure}
\begin{center}
\includegraphics[width=12cm]{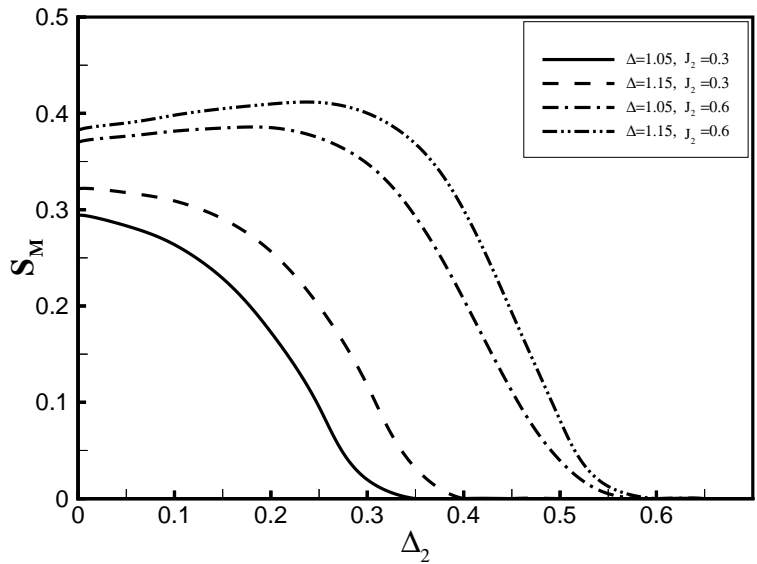}
\caption{The staggered Magnetisation in z-direction for
different
values  of $\Delta$ and $J_2$ versus $\Delta_2$. 
The order parameter ($S_M$) goes continuously to zero
which represents the transition from N\'{e}el to dimer
phase.} \label{fig4}
\end{center}
\end{figure}

In the region of the antiferromagnetic anisotropy ($\Delta>1$)
where the N\'{e}el phase exists, it is destabilized by sufficiently
strong competing next-nearest neighbour couplings,
$J_{2}^{c}$ and $\Delta_2^c$. The N\'{e}el phase appears
just by crossing the $\Delta=1$ plane at $\Delta_2=0$ and $J_2=0$.
In the $\Delta_2=0$ plane and for $\Delta >1$ the model will
pass through a phase transition from  N\'{e}el to dimer phase
for $J_2 > J_2^c(\Delta)$. This boundary is the intersection 
of the semi-circular lines and the $\Delta_2=0$ plane which happens 
around $J_2\sim 1.5$ (see Fig.(\ref{fig2})). 
The N\'{e}el ordered is also broken by increasing the
anisotropy of the n-n-n interaction. This is a phase transition to the dimer phase
which is shown in Fig.(\ref{fig2}) by the semi-circular lines where the dimer
phase exists above them.
Thus, both the n-n-n exchange ($J_2$) and n-n-n anisotropy ($\Delta_2$) drive
the model from the N\'{e}el ordered phase to the dimer one.
In this respect, the boundary between the N\'{e}el and the dimer
phase looks like an arcade in the phase diagram. 

We have probed the boundary of the
N\'{e}el-dimer transition by calculating the staggered
magnetisation in the $z$-direction as an order parameter
(Fig.(\ref{fig4})),
\be
S_M=\frac{1}{N}\sum_{i=1}^{N} (-1)^i \langle \sigma_i^z \rangle.
\label{sm}
\ee
The staggered magnetisation
($S_M$) is zero in the dimer phase and has a
nonzero value in the N\'{e}el phase. Thus the staggered magnetisation
is the proper order parameter to represent the 
N\'{e}el-dimer transition. We have plotted $S_M$ versus $\Delta_2$
for different values of $\Delta=1.05, 1.15$ and $J_2=0.3, 0.6$ in Fig.(\ref{fig4}).
The staggered magnetization goes to zero continuously at a critical value
of $\Delta_2^c(\Delta, J_2)$ which shows the destruction of N\'{e}el order.

A similar calculation has been done to trace the phase transition between
the N\'{e}el and the spin-fluid phase.
The crossing boundary of the fluid and N\'{e}el phases always
stays at $\Delta=1$. The reason is related to the $SU(2)$ symmetry of the
Hamiltonian on the line $\Delta=\delta=1$. The QRG approach preserves this
symmetry which can be seen by the irrelevant direction of $\Delta_2=J_2$ in
the $\Delta=1$ plane toward the tri-critical point.

The significant result of our calculations occurs at the
isotropic plane, $\Delta=1$. 
For small $J_{2}$ the RG equations show running
of $J_{2}$ to zero except at the isotropic plane ($\Delta=1$).
This means, if we start with the $XXX$ model ($J_2=0$), the next-nearest
neighbour interaction ($J_{2}$) is generated and runs to the
tri-critical point, $\Delta_2^*=J_{2}^*=0.155, \Delta^*=1$. It might be interesting to
mention that this fixed point corresponds to the reported
experimental value for $CuGeO_{3}$ ($J_{2}=0.13$) \cite{Mizuno}.
It is worth mentioning that for the isotropic case $\Delta=\delta=1$, 
if the projection operator for the
second order RG, ($1-P_0$), does not contain all of the remaining
states of the block we get $J_{2}^*=0.0761$ \cite{miguel-2}. This
shows the increase of $J_{2}^*$ when we have considered the
correlation effects more precisely. Thus, we expect a nonzero
value for $J_{2}^*$ if we go further to consider the higher order
RG approximation. This is also true for $J_{2}=0$ as the initial
value, the corresponding fixed point is nonzero. So, the
manageable fixed point, the fixed point which enables us to get
the physical quantity, represents the next-nearest neighbour
interaction for the isotropic Heisenberg chain. At this fixed
point the correlation length is zero and the system behaves as a
classical one. It represents a frustrated chain of classical spins
which does not posses any ordering. It might be an explanation why
the nearest neighbour one-dimensional antiferromagnetic Heisenberg chain does not
show a true long range order.

The other remarkable result of our RG flow is the observation of
the specific lines in the spin-fluid and N\'{e}el phases which has
been shown by dashed lines in Fig.(\ref{fig3}). We have linearised
the RG flow at ($\Delta_2^*=J_{2}^*\simeq0.155,\Delta^*=1$), and found one
relevant and two irrelevant directions. 
The eigenvalues of the matrix of linearised flow are $\lambda_1=1.24$, 
$\lambda_2=0.46$ and $\lambda_3=0.41$. The corresponding eigenvectors
in the $|\Delta, J_2, \Delta_2\rangle$ coordinates are 
$|\lambda_1\rangle=|-0.98, 0.02, -0.20\rangle$,
$|\lambda_2\rangle=|-0.88, 0.15, -0.44\rangle$
and
$|\lambda_3\rangle=|0, 0.707, 0.707\rangle$.
The direction of the dashed lines in Fig.(\ref{fig3}) close
to the tri-critical point corresponds to $|\lambda_1\rangle$.
The two dashed lines 
start at $\Delta^*=1,J_{2}^*$, one of them ends at the XY fixed point ($\Delta=0, J_{2}=0$)
and the other ends in the Ising fixed point ($\Delta=\infty, J_{2}=0$) . Although these lines
separate two different RG flows below and above the dash lines 
the flows represent a unique phase which is
the spin-fluid for $\Delta < 1$ and the N\'{e}el phase for $\Delta >1$. 
These lines are supposed
to be the Gaussian fixed points which have been reported in
Ref.\cite{Nomura}. In the spin-fluid and N\'{e}el phases, large
distance physics is described by the Gaussian model where the
system possesses the conformal symmetry \cite{Francesco}.

We have also calculated some critical exponents at the tri-critical point.
In this respect, we have obtained the dynamical exponent, the exponent of
order parameter and the diverging exponent of the correlation length.
This corresponds to reaching the
tri-critical point from the N\'{e}el phase by approaching $\Delta \rightarrow 1$.
The scaling of gap exponent is $z \simeq 0.83$, the staggered magnetization goes
to zero like $S_M \sim |\Delta -1|^{\beta}$ with $\beta\simeq0.82$ and the correlation
length diverges at $\Delta=1$ with exponent $\nu\simeq2.27$.
The detail of a similar calculation can be found in Ref.\cite{miguel1}.


\section{summary and discussions}
We have presented the second order RG approximation to obtain the
phase diagram, staggered magnetisation and ground state energy of
the XXZ chain with next-nearest neighbour interaction. We have
observed that the second order RG equations must be considered to get a
self similar Hamiltonian upon renormalisation processes.
The ground state energy density of the nearest neighbour $XXZ$
($J_2=0$) model in the limit $N \longrightarrow\infty$ is then
given by,
$e^{BRG}_{\infty}=\sum_{m=0}^{\infty}\frac{1}{3^{m+1}}e_{B}(J^{(m)},\Delta^{(m)})$
where initially $J^{(0)}=J$ and $\Delta^{(0)}=\Delta$. We have
considered the nearest neighbour case to be able to compare with
the exact results. We have shown this quantity in Table.(1) where
the results of the first and second order RG presented for $0\le
\Delta <0.9$. The relative error has been derived by comparison
with the exact result \cite{Cloizeaux}. The second order
correction improves the relative error at least to 65 percent.
However the improvement of the 2nd order approximation is not monotonic for the
whole range of $\Delta$.

\begin{table}[t]
\begin{center}
\label{table1}
\caption{The first
($e^{(1)}_{BRG}=\frac{E^{(1)}_{BRG}}{JN}$) and second
($e^{(2)}_{BRG}=\frac{E^{(2)}_{BRG}}{JN}$)
order approximations of the ground state energy per site
are compared with the exact ($e^{exact}=\frac{E^{exact}}{JN}$) value
\cite{Cloizeaux}.
The relative errors, $\epsilon^{(1)}=\frac{E^{(1)}_{BRG}-E^{exact}}{E^{exact}}$ and
$\epsilon^{(2)}=\frac{E^{(2)}_{BRG}-E^{exact}}{E^{exact}}$ have
also been shown.\\}
\begin{tabular}{cccccc}
\hline $\Delta$&$e^{exact}$&$e^{(1)}_{BRG}$&$e^{(2)}_{BRG}$&
$\epsilon^{(1)}$&$\epsilon^{(2)}$ \\
\hline
$0.0$  &  $-0.3183$  &  $-0.2828$  &  $-.03229$  &  $11.1\%$  &  $2.3\%$  \\
$0.1$  &  $-0.3286$  &  $-0.2921$  &  $-0.3319$  &  $11.1\%$  &  $1\%$    \\
$0.2$  &  $-0.3395$  &  $-0.3016$  &  $-0.3390$  &  $11.1\%$  &  $0.14\%$ \\
$0.3$  &  $-0.3509$  &  $-0.3115$  &  $-0.3470$  &  $11.2\%$  &  $1.1\%$  \\
$0.4$  &  $-0.3627$  &  $-0.3217$  &  $-0.3559$  &  $11.3\%$  &  $1.8\%$  \\
$0.5$  &  $-0.3750$  &  $-0.3322$  &  $-0.3656$  &  $11.4\%$  &  $2.5\%$  \\
$0.6$  &  $-0.3877$  &  $-0.3432$  &  $-0.3759$  &  $11.4\%$  &  $3.04\%$ \\
$0.7$  &  $-0.4009$  &  $-0.3545$  &  $-0.3870$  &  $11.5\%$  &  $3.4\%$  \\
$0.8$  &  $-0.4145$  &  $-0.3663$  &  $-0.3988$  &  $11.6\%$  &  $3.7\%$  \\
$0.9$  &  $-0.4286$  &  $-0.3785$  &  $-0.4112$  &  $11.6\%$  &  $4.05\%$ \\
\end{tabular}\\
\end{center}
\end{table}


An interesting result of the QRG scheme is the case of isotropic
Heisenberg model ($J_2=0, \Delta=1$) which runs to the frustrated
next-nearest neighbour fixed point under the second order RG
equations. This can be viewed as an explanation for the absence of
long range order in the nearest neighbour antiferromagnetic
Heisenberg model. Moreover, our primary calculation to take into account the
higher order RG equations shows that the effective (renormalised)
Hamiltonian will consist of the long range interactions. Taking into
account the third order correction will result in an infinite range
interaction of XXZ type. We would also like to mention a comment
on the implementation of the 2nd order RG approach. The
calculation is straightforward (not necessarily simple) whenever
the zeroth order projection operator ($P_{0}$) is composed of
degenerate states. This makes possible to obtain the effective
Hamiltonian in the analytic form. We have tried to get the 2nd
order corrections for the $S=\frac{1}{2}$ $XYZ$ model in the
transverse field where the block ground-state is not degenerate.
We were not able to obtain the analytic 2nd order corrections.
This is important because the advantage of the QRG is the analytic
form of results which helps to get the phase diagram 
even though it may not be quantitatively very accurate.
If we are forced to do the RG approach in a numerical
way then it is appropriate to implement the Density Matrix
Renormalisation Group (DMRG) method. In that case, we loose the renormalisation
group features of the running of the coupling constants but get the 
accurate numerical values.

\section{Acknowledgment}
The authors would like to thank P. Thalmeier for reading the manuscript and his comments.


\end{document}